\documentclass{aa}
\input epsf
\newcommand{\PSbox}[3]{\mbox{\includegraphics{#1}\hspace{#2}\rule{0pt}{#3}}}

\begin{document}
   \thesaurus{}

   \title{Gamma-ray Burst Positions from the ASM on RXTE}

   \author{Hale V. Bradt \& Donald A. Smith}

   \offprints{H. Bradt}

   \institute{Department of Physics \& Center for Space Research,
              Massachusetts Institute of Technology,
              Cambridge, MA 02139-4307 USA}

   \date{}
\authorrunning{H. V. Bradt \& D. A. Smith}
\titlerunning{GRB from the ASM on RXTE}
   \maketitle

   \begin{abstract}
\vspace{-5mm}

The RXTE/ASM has detected and positioned 14 confirmed GRB bursts (at
this writing, Jan. 1999) including six whose positions were
comunicated to the community 2 to 32 hours after the burst. Two of
these latter bursts led to measurements of optical red shifts but one,
despite an easily detected x-ray afterglow, produced no detectable
optical or radio afterglow.

   \vspace{-2mm}

   \end{abstract}

\vspace{-5mm}

\section{ASM capabilities}

The All-Sky Monitor on RXTE (\cite{Lev96}) has the capability to
locate Gamma-Ray Bursts (GRBs) to within a few arcminutes in two
dimensions. This can occur if the burst falls within the
$\sim10^\circ$ by $\sim50^\circ$ parallelogram on the sky which is
viewed simultaneously by the two azimuthal shadow cameras of the ASM
(see \cite{Lev96}). The solid angle for such detections is actually
somewhat larger in some cases because the burst intensity (or that of
its immediate afterglow) may remain above ASM threshold as the ASM
steps to its next celestial position, {\it i.e.} to its next
``dwell''. This will sometimes bring the burst into the field of view
of a camera that had not yet detected it. The ASM takes data for
$\sim40\%$ of the orbital time.

It is more probable that the burst will fall within the FOV of only
one of the three shadow cameras, each with a field of view of
$\sim10^\circ \times \sim100^\circ$. In the case of a detection in 
only one collimator, the error
region will be a few arcminutes wide and a few degrees long.

The RXTE does not carry a dedicated GRB detector, and the ASM detects
only the x-ray portion of GRB spectra, 1.5--12 keV. It is therefore
difficult to distinguish a rapid x-ray transient from a GRB. Our
original method for securely identifying a new transient was to wait
until several sightings confirmed the detection. This would
distinguish a genuine transient from background events. In the case of
GRB, one may have only one sighting. This requires one to impose a
higher threshold for detection.

In addition to ``position data'' that yields intensities and locations
of a source based on 90-s integrations, the ASM records ``time-series
data'' in 1/8 s time bins for each of the three energy channels of
each of the three detectors. If a new source is detected in the
position data of a given ASM dwell, the presence of rapid variability
in the corresponding time-series data improves the likelihood that the
source is a GRB.

\vspace{-5mm}

\section{ASM burst detections}

The RXTE/ASM has detected and positioned 14 GRB since Feb. 1996. Each
of the 14 has been confirmed as a burst with detections from GRB
detectors on one or more other satellites. Seven of these were
detected in two of the ASM shadow cameras, thus yielding positions
accurate to a few arcmintues in two dimensions. Of the 14, eight were
located in searches of archived data.  One of these (GRB 961216) was
near the edge of the ASM camera field of view and thus led to a large
(and uncertain) error region. The other 7 archival detections and
positions are reported in Smith et al. (1999) together with detections
and positions of these events from other satellites. They are GRB 960416, 
960529, 960727, 961002, 961019, 961029, and 961230. 

Beginning in August 1997, bursts detected by the ASM were analyzed and
positions were reported in near real time. Six events have been so
reported at times ranging from 2 to 32 hours after the burst. The
error regions reported by us and by others for these six bursts are
shown in Figure 1. Notable among these are (1) GRB 970828
(\cite{Rem97}) which had an easily detected x-ray afterglow but no
optical or radio signal (\cite{Gro98}), (2) GRB980703 (\cite{Lev98})
which led to a radio/optical transient with a redshift z = 0.9660
(\cite{Djo98}), and (3) GRB981220 (\cite{Smi98b}) which led to the
discovery of a highly variable radio source as a possible counterpart
candidate (\cite{Gal98}, \cite{Fra98}).  The radio source was
associated with a faint galaxy at R = 26.4 (\cite{Blo99}). The GRB
971214 led to a redshift of 

\onecolumn
\PSbox{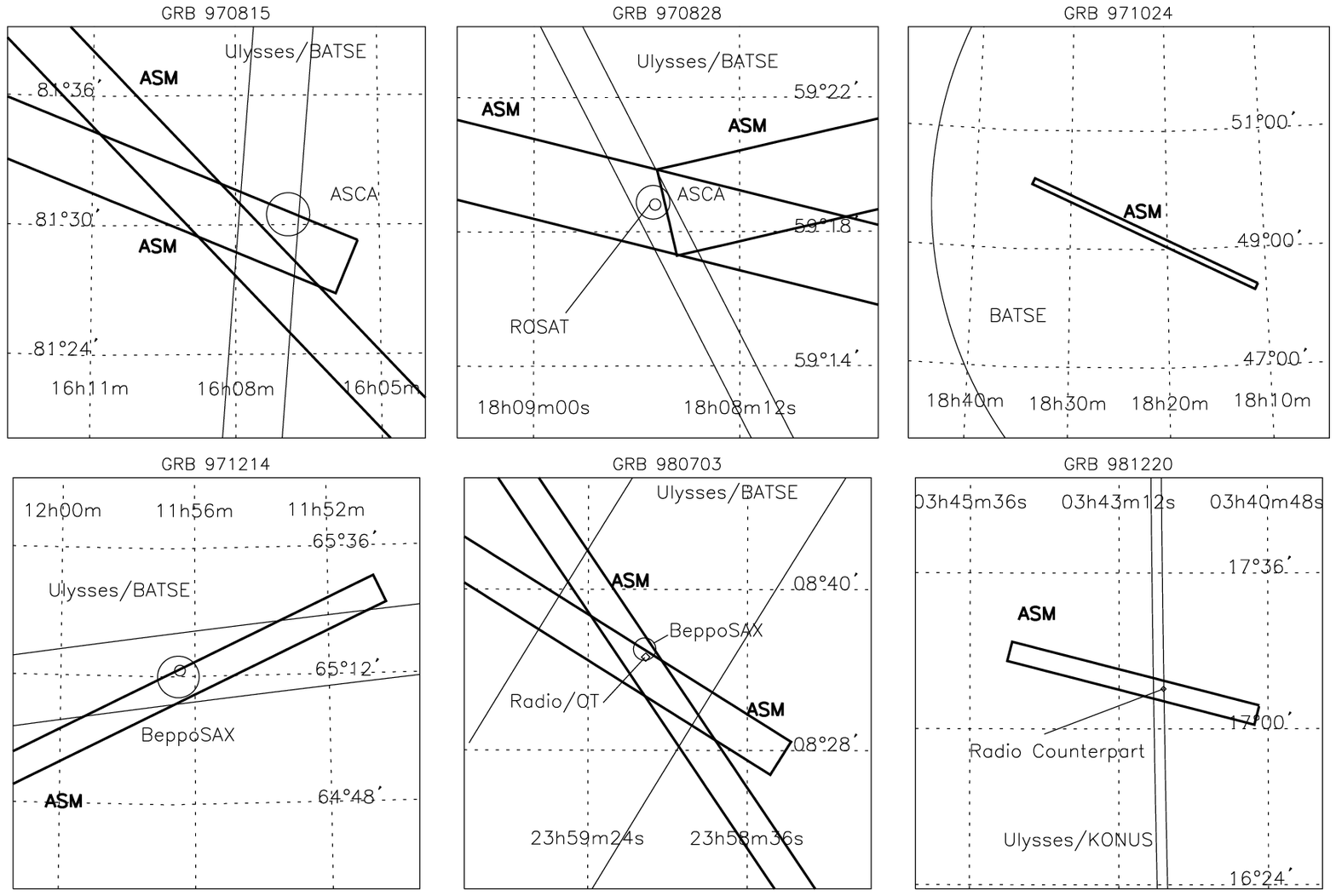 hoffset=18 voffset=0 hscale=87
vscale=87}{18cm}{10cm}{\\ Figure 1. ASM Positions of the six
bursts reported in near real time by the ASM group together with
refined BATSE and Interplanetary Network positions from other groups
as reported or referenced in Smith et al. 1999. }

\hbox{\parbox{8.8cm}{ 
$z$ = 3.42 but this result came primarily through the
precise two-dimensional positions from SAX (\cite{Hei97};
\cite{Ant97}).  The ASM results and associated studies are reported or
referenced in Smith {\it et al.} 1999.

\vspace{-5mm}

\section{New rapid GRB position GCN alerts}

The capability of the ASM system to produce rapid positions of GRB is
improving steadily. We have just initiated (Jan. 1999) automatic
transmissions to the GCN when the ASM detects a new transient in two
different collimators with sufficient signal strength to rule out
noise signals with high confidence. These events can be either a GRB
or a new transient, and they will have position precision of a few
arcminutes in two dimensions. The trigger criteria would have picked
up 6 of the 7 two-collimator burst detections recorded since 1996. In
addition, it would also have picked up five new x-ray transients in
the past year. For example, a preliminary version of this system
detected the bright black hole transient XTE J1550--564 at an
intensity of only 70 mCrab, very early in its rise from anonymity on 7
Sep 1998 (\cite{Smi98a}). The position notices are now automatically
transmitted to the GCN within 5--60 minutes of the burst arrival time
under normal operating conditions. We expect that about 1 to 2 GRB
events per year will satisfy the trigger criteria.

}
\hspace{0.3cm}
\parbox{8.8cm}{

\begin{acknowledgements}

The entire ASM team at M.I.T. has contributed to this work. We are grateful 
for the GRB groups of BATSE, Ulysses, and Konus for providing refined GRB 
positions for inclusion in Smith et al. 1999, some of which are reproduced 
herein in Fig. 1. Support for the ASM work was provided in part by NASA 
Contract NAS5--30612.

\end{acknowledgements}

}}


\begin{thebibliography}{}
\bibitem[Antonelli {\it et al.} 1997]{Ant97}Antonelli, A. {\it et al.} 
1997, IAU Circ. 6792
\bibitem[Bloom {\it et al.} 1999]{Blo99} Bloom, J., Djorgovski, S., 
Kulkarni, S., Brauher, J., Frail, D., Goodrich, R., \& Chaffee, F. 1999, 
GCN Circ. 196
\bibitem[Djorgovski {\it et al.} 1998]{Djo98} Djorgovski, S., Kulkarni, S., 
Bloom, J., Goodrich, R., Frail, D., Piro, L., \& Palazzi, E. 1998, ApJ, 
508, L17
\bibitem[Frail \& Kulkarni 1998]{Fra98} Frail, D. \& Kulkarni, S. 1998, GCN 
Circ. 170
\bibitem[Galama {\it et al.} 1998]{Gal98} Galama, T., Vreeswijk, P., van 
Paradijs, J., Kouveliotou, C., Strom, R., \& de Bruyn, G. 1998, GCN Circ. 
168
\bibitem[Groot {\it et al.} 1998]{Gro98} Groot, P., {\it et al.} 1998, ApJ, 
493, L27
\bibitem[Heise {\it et al.} 1997]{Hei97}Heise, J. {\it et al.} 1997, IAU 
Circ. 6787
\bibitem[Levine {\it et al.} 1996]{Lev96}Levine, A., Bradt, H., Cui, W., 
Jernigan, J.G.,  Morgan, E., Remillard, R., Shirey, R., Smith, D.A. 1996, 
ApJ, 469, L33
\bibitem[Levine, Morgan \& Muno 1998]{Lev98}Levine, A., Morgan, E. \& Muno, 
M. 1998, IAU Circ. 6727
\bibitem[Remillard {\it et al.} 1997]{Rem97} Remillard, R., Wood, A., Smith, D. A., \& Levine, A. 1997, IAU Circ. 6726
\bibitem[Smith 1998a]{Smi98a} Smith, D. A. 1998a, IAU Circ. 7008
\bibitem[Smith 1998b]{Smi98b} Smith, D. A. 1998b, GCN Circ. 159
\bibitem[Smith {\it et al.} 1999]{Smi99} Smith, D. A., Levine, A. M., Bradt, H. V., Remillard, R., Jernigan, J. G., Hurley, K. C., Wen, L., Briggs, M., Cline, T., Mazets, E., Golenetskii, S., \& Frederics, D. 1999, ApJ, submitted.

\end{thebibliography}
\end{document}